\documentclass[manuscript]{aastex}
\usepackage{url}
\usepackage{lscape} 

\RequirePackage{color}

\begin{document}

\title{K-band spectra of selected post-AGB candidates}
\slugcomment{Not to appear in Nonlearned J., 45.}
\shorttitle{K-band spectra}
\shortauthors{Parthasarathy et al.}

\author{M.\,Parthasarathy\altaffilmark{}} \and \author{C.\,Muthumariappan\altaffilmark{}} \and
\author{S. Muneer\altaffilmark{}}



\altaffiltext{}{Indian Institute of Astrophysics, \\Koramangala, Bangalore, 560 034 India}
\email{m-partha@hotmail.com}

\begin{abstract}
We present medium resolution (1000) K-band spectra of 12 post-AGB candidates and related stars.
 For several objects in our sample, these spectra were obtained for the first time. The Br$\gamma$
 line in emission is detected in seven objects indicating the onset of photo-ionization in 
 these objects. Four objects show the presence of He\,{\sc i} line. We detect H$_{2}$ emission line 
 in the spectra of IRAS\,06556+1623, IRAS\,22023+5249, IRAS\,18062+2410 and IRAS\,20462+3416. 
 H$_{2}$ emission line ratio 1-0 S(1)/2-1 S(1) indicate that H$_{2}$ is radiatively excited due to the UV 
 radiation of hot post-AGB central stars.
  When compared with the recent observations
 by other investigators, the Br$\gamma$ and H$_{2}$ emission fluxes varied in some of the objects.
 The hot post-AGB stars IRAS\,22495+5134, IRAS\,22023+5249, IRAS\,18062+2410 and IRAS\,20462+3416 
 seem to be evolving rapidly to young low excitation planetary nebula phase. The spectra of the
objects presented in this paper may be useful for future observers as some of these stars show
spectral variation and since the post-AGB evolution of some of these stars is relatively rapid. 
\end{abstract}



\section{Introduction}

The early stage of post Asymptotic Giant Branch (post-AGB) stars is marked by the  
 spectral types typically from K\,I to F\,I type supergiants. Their ejected envelopes expand and can be seen in the 
 scattered stellar light and in molecular lines such as  H$_{2}$ and CO. They mimic the spectra of supergiants because of their extended atmospheres. 
 As the cool post-AGB objects evolve they will be hot post-AGB stars of the spectral types A\,I to OB\,I supergiants. 
  They are rapidly growing hotter 
 due to their gravitational contraction as they evolve on the post-AGB evolutionary track towards young planetary nebulae (PNe) stage \citep{1986A&A...154L..16P,1989A&A...225..521P}.  
 The UV radiation of the post-AGB star causes photoionization of the envelope and the spectrum shows recombination lines and collisionally excited lines.
 \citet*{partha93a} listed several cool and hot
post-AGB candidates based on their IRAS colors and flux distribution. Among those post-AGB objects 
 which have far-IR colours similar to PNe, several proto-PNe with hot (O\,I to B\,I type) stars were 
 detected \citep{partha00a}. The UV spectra of some of these objects show violet 
 shifted stellar wind P-Cygni profiles indicating hot and fast radiatively driven stellar winds. Multi-wavelength studies of these objects enabled us to further 
 understand the evolution of their circumstellar envelopes and that of the central stars.

 Near-Infrared (Near-IR), medium resolution ($\sim$\,1000) spectroscopy in K-band (from 2 -- 2.4 
 $\mu$m) is quite useful and a convenient tool to study these objects as it covers a range of
 emission spectra to trace their molecular as well as ionized envelopes. The excitation 
 and ro-vibrational transitions of molecular hydrogen (at 2.1218-, 2.2477-, 2.2233-, 2.2014 and
 2.1542 $\mu$m) and the recombination lines of hydrogen (Br$\gamma$ line at 2.166 $\mu$m) and of
 helium (He\,{\sc i} line at 2.058 $\mu$m) are present in near-IR spectrum. \citet*{gledhil15} give the spectral
features and their wavelengths in the K-band spectra of several hot post-AGB stars. They carried out integral field spectroscopy
of hot post-AGB stars with near-infrared Intgral Field Spectrometer (NIFS) instrument on Gemini North. It provided much higher resolution 
and much better sensitivity, and spectral images. They made observations in 2007 and have three sources in common with the present study. Their work
revealed spatial extent of the emission line regions in those three objects (see Section 3.0, Figure 1, Table 1).

Emission line ratios of 
 different transitions of H$_{2}$ (1-0 S(1)/2-1 S(1) and 1-0 S(1)/3-2 S(3)) give a clue on the 
 dominant excitation mechanism of H$_{2}$, i.e. due to fluorescence by the UV radiation from the
 central star or thermally by shocks. While the emission lines of H$_{2}$ (and also CO lines at 
 2.2935 $\mu$m and 2.322 $\mu$m) help us to trace the molecular gas, Br$\gamma$ and He\,{\sc i}  emission 
 lines come from the recombination spectra of ionized hydrogen and singly ionized helium respectively
 and hence they trace the ionized gas in the nebula. Br$\gamma$ emission traces the ionized hydrogen
 gas better than the optical lines as the optical lines suffer high extinction due to the presence 
 of thick circumstellar dust envelopes. Near-infrared spectra of post-AGB stars (proto-planetary nebulae PPNe)
was studied by \citet*{hriv94}, \citet*{1995A&A...299...69O}, \citet*{garcia02}, \citet*{kelly15} and \citet*{gledhil15}. The observations 
reported here were made in October 1999 and are similar in sensitivity and resolution to observations made by  \citet*{garcia02}. 
 We present here K-band medium resolution (1000) spectra of 12 selected  post-AGB candidates (nine are hot stars and three are cool stars). 

\section{Observations}

The observations were made with the 188-cm telescope of the Okayama Astrophysical Observatory (OAO)
 of the National Astronomical Observatory of Japan (NAOJ) and the Okayama Astrophysical System for 
 Infrared Imaging and Spectroscopy OASIS \citep{oka2000} was used  in obtaining the spectra. 
 The OASIS was used in spectroscopic mode and K-band spectra covering the wavelength range 2.0 --
  2.4\,$\mu$m were obtained during 1999 October 22 to 25. We used 300 lines per mm grating 
 and a slit width of 2.4 seconds of arc. The slit position angle was fixed to East-West. The spectra 
 (Figure 1) have a resolution of about 1000. We obtained spectra of rapidly rotating B and A type 
 stars with no stellar emission lines in their K-band spectra to remove atmospheric absorption lines
 in the K-band spectra of stars presented in the paper.

  Standard process of data reduction was performed with IRAF. The wavelength calibration was 
 accomplished using the atmospheric OH lines \citep{oli92}. The sky background was removed
 by fitting the spectrum of the sky adjacent to the program star. The final spectra of the program 
 stars are shown in Figure 1. The line fluxes are measured and are listed in Table 1 along with the
 emission line ratios of H$_{2}$ 1-0 S(1)/2-1 S(1). We detected He\,{\sc i} line at 2.058\,$\mu$m for some sources
 however, full line profile is not recorded to measure the flux. Hence in Table 1 we only indicate 
 if this line is present in the spectrum. 

 \section{Notes on individual objects}

 We discuss here the K-band spectral properties of each post-AGB candidate based on Figure 1 
 and Table 1. The spectral-types of these 12 objects can be found in SIMBAD (see Table 1). 
 
 \subsection{IRAS\,22495+5134 (M2-54, LS III +51 42)} 

  Optical variation in the
  brightness and spectrum of IRAS\,22495+5134 were studied by \citet[see also \citealt{1999A&AS..135..493H}]{arkh13}.
  It shows small amplitude variability in brightness and also variability in the spectrum.
   We find from our spectrum of this star strong Br$\gamma$ emission profile 
  showing the developed hydrogen photo-ionization region in the nebula. H$_{2}$ lines are not
  detected in our spectrum. It is a young compact planetary nebula. The UV spectrum of IUE and the circumstellar dust characteristics of
  this object were analyzed by \citet{gauba03,gauba04}. From the UV spectrum they derive 
  that the central star has a spectral type of O9. This object seem to be evolving rapidly 
  to the planetary nebula phase. H$_{2}$ emission is not detected in the K-band spectrum of this source (Figure 1). The central star is
  hot enough for radiative H$_{2}$ excitation, but previous studies argued that H$_{2}$ is expected to be
  disassociated for hot post-AGB stars unless they have a thick torus region that provides shielding.   \\
 
 \subsection{IRAS\,06556+1623 (HD\,51585)} 

  IRAS\,06556+1623 is a high Galactic latitude (b = 8$^{o}$.9), with Be I/BQ[] spectral type. 
  \citet{1989A&A...225..521P} and  \citet{partha93a} classified this as a hot post-AGB star based on IRAS colours and flux 
  distribution. \citet{arkh06} and \citet{1996A&AS..117..281J} detected optical spectroscopic and photometric variability 
  of this star. We present the K-band spectrum of this object and it shows a strong Br$\gamma$
  in emission which is due to a developed hydrogen photo-ionized region. In short ward of Br$\gamma$,
  there seem to be weak emission lines present in the spectrum which could be due to Fe. H$_{2}$ 
  lines are seen in the spectrum, which are much weaker than Br$\gamma$ line (Figure 1). We find that H$_{2}$ 
  line flux ratios of 1-0 S(1)/2-1 S(1) = 4.73 (Table 1). For a purely UV pumped H$_{2}$ excitation, this ratio 
  is about 2 \citep{black76} and it can have a value between 4 to 10 for a shock excitation 
  \citep{smi95}. However, for dense gas exposed to intense UV radiation, collisional heating causes 
  this ratio little larger than expected from pure UV pumping. 
  \citet{gledhil15} stated in their paper that shocks were not evident 
  in most of their hot post-AGB sources and argued for UV excitation with collisional heating. They also
  showed line ratio maps that illustrated how the 1-0/2-1 ratio can vary across a source.  \citet{garcia02} included this star in their IR spectroscopic survey of post-AGB stars 
  and found this ratio to be larger than 1.7. Our K-band spectrum also shows He\,{\sc i} line at 2.058 $\mu$m
  indicating singly ionized helium region and a hotter star which is consistent with the presence of a
  photoionized nebula. In the optical spectrum of this star, in addition to
  nebular emission lines, several Fe lines were observed by \citet{arkh06}. 
  
  Figure 2 shows a medium resolution (1000) optical spectrum of IRAS 06556+1623. The optical spectrum of the object was obtained on 
  January 28, 2003 using the OMR spectrograph at the cassegrain focus of the 2.3-m Vainu Bappu Telescope (VBT) at 
  Kavalur. A dispersion of 2.6 \AA\ per pixel is achieved  using a 600 lmm$^{-1}$ grating. The following lines are seen in the 
  spectrum: 
  H {\sc i} (4340 \AA), 
  [O\,{\sc iii}] (4363 \AA), 
  He {\sc i} (4388 \AA), 
  He {\sc i}  (4472 \AA), 
  H {\sc i} (4861 \AA), 
  He {\sc i} (4922 \AA), 
  [O {\sc iii}] (5007 \AA), 
  He {\sc i} (5876 \AA), 
  [O {\sc i}] (6300 \AA), 
  [O {\sc i}] (6364 \AA), 
  H {\sc i} (6563 \AA), 
  He {\sc i} (6678 \AA), 
  He {\sc i} (7065 \AA) and 
  [O {\sc ii}] (7325 \AA) (see Figure 2). The presence of  [O {\sc iii}] 5007 \AA\ 
  forbidden line in the optical spectrum (\citealt{1996A&AS..117..281J}) indicates photoionization has started. The presence of He\,{\sc i} emission lines shows 
  that the central star is of high T$_{eff}$. This star may have a compact nebula and disc around it. \citet{arkh06} find small amplitude 
  variation in brightness and variation in the strength of emission lines (see also \citealt{1996A&AS..117..281J}). \\
 
 \subsection{IRAS\,18237-0715 (MWC 930)}
  
   \citet{vijap98} presented the low resolution blue spectrum of IRAS\,18237-0715 and concluded
 that it may be a post-AGB star or a Luminous Blue Variable (LBV, see also \citealt{partha00a}). 
 The stellar spectral type is B5/9 Iaeq. A detailed optical spectroscopic study of this star 
 was made by \citet{miro05} and they found it to be a new LBV. We present the K-band spectrum 
 of this object for the first time and our spectrum shows P-Cygni type Br$\gamma$ emission profile
 showing ongoing mass-loss and also few  emission lines short ward of Br$\gamma$. The Br$\gamma$ emission
 indicates the presence of hydrogen photo-ionization region in this object. We also detected He\,{\sc i} emission line 
 at 2.058 $\mu$m. Our spectrum of this object does not show 
 H$_{2}$ lines (Figure 1). The non-detection of H$_{2}$ could be a sensitivity issue. High-resolution K-band spectrum with better sensitivity 
 may reveal H$_{2}$. \\  
 
 \subsection{IRAS\,22023+5249 (LS III + 52 24)}

  IRAS\,22023+5249 is a high velocity hot post-AGB star \citep{sark12}. From 
  their 2\,$\mu$m spectral survey, \citet{kelly15} find H$_{2}$ emission from this object. 
  From our K-band spectrum, we also find H$_{2}$ emission lines
  at 2.12-, 2.22- and 2.247 $\mu$m (Figure 1). We also find Br$\gamma$ emission from this object indicating the presence  
  of low excitation nebula. The Br$\gamma$ flux is comparable to the flux of H$_{2}$ at
  1-0 S(1) transition. However no He\,{\sc i} line is detected. We find that the H$_{2}$ 1-0 S(1)/2-1 S(1) line
  flux ratio is 3.02 indicating the excitation mechanism of H$_{2}$  
  is due to UV fluorescence. \citet{kelly15} find this ratio to be 2.7 which is in good agreement with our value. \citet{gledhil15}  
  made a detailed study of the 2 $\mu$m to 2.4 $\mu$m spectrum of this source. They show that the
  H$_{2}$ 1-0 S(1)/2-1 S(1) ratio has value of 3.41 and also suggested the excitation of H$_{2}$ can be
  contributed both by shocks and UV pumping. The IUE UV spectrum and infrared data of the circumstellar dust shell 
  of IRAS\,22023+5249 were studied by \citet{gauba03,gauba04}. They suggested the presence of a dusty disk
  around the star. They also find evidence for the presence of hot stellar wind from the IUE spectrum. This object seems to be evolving rapidly towards 
  planetary nebula phase. The star shows small amplitude variability in brightness and also variations in the spectral lines \citep{arkh13}. 
  \citet{gledhil15} show a map of H$_{2}$ ratios (see Figure 9 in their paper), which shows ratios of 5 $-$ 8 in bubble-like lobes and extended H$_{2}$
  with ratios of three and lower. \\

 \subsection{IRAS\,18062+2410 (SAO 85766)}

  IRAS\,18062+2410 is a high Galactic latitude (b=19$^{o}$.8) hot post-AGB star. It is a rapidly
  evolving hot post-AGB star \citep{partha00b} similar to SAO 244567 \citep{partha93,partha95} 
   with Be spectral type. Our K-band spectrum appear to some extent similar to that of 
  IRAS\,22023+5249. We find H$_{2}$ emission lines at 2.12-, 2.22- and 2.247 $\mu$m in our K-band
  spectrum (Figure 1). The H$_{2}$ line flux ratio 1-0 S(1)/2-1 S(1) = 4.89. \citet{kelly15} find 
  from their 2 micron spectral survey that this ratio has a value of 3.8 and also proposed that 
  H$_{2}$ excitation has both thermal and radiative origin. \citet[and references therein]{gledhil15} 
  have made a detailed study of IRAS\,18062+2410 using their 2\,$\mu$m to 2.4\,$\mu$m spectrum and
  show that H$_{2}$ line ratio S (1)/2-1 S(1) = 6.2. Br$\gamma$ emission shows a 
  P-Cygni profile showing ongoing post-AGB mass-loss, and the presence of this
  line indicates the onset of photo-ionized hydrogen region. The Br$\gamma$ flux is smaller than the flux of
  H$_{2}$ at 1-0 S(1) transition. The strength of Br$\gamma$ is variable with time \citep{garcia02}.  
  \citet{gledhil15} conclude that there is little evidence for shock structures in this source and that the high 1-0/2-1 ratio came from
  regions where the gas density and UV intensity are high. \\

 \subsection{IRAS\,18313-1738 (MWC 939)}

  \citet{vijap98}  and also \citet{partha00a} presented the blue spectrum of
  IRAS\,18313-1738. Its spectral type is Be IV and the object is most likely  a hot post-AGB star \citep{partha00a} or 
  a hot sub-dwarf. 
  We present the K-band 
  spectrum of this object for the first time and it also has a hydrogen photo-ionized region as shown 
  by the presence of Br$\gamma$ emission line. He\,{\sc i} line is not present in the K-band spectrum. We do not 
  detect H$_{2}$ lines in our spectrum. The non-detection could be a sensitivity and resolution issue.  \\ 

 \subsection{IRAS\,20462+3416 (LS II +34 26)} 
 
  \citet{partha93b} discovered ~ IRAS  20462+3416 as a hot post-AGB star.  
   From our 
  K-band spectrum we find the presence of H$_{2}$ lines at 2.12-, 2.22- and 2.247 $\mu$m. We also 
  detected Br$\gamma$ emission (Figure 1) which shows the presence of hydrogen ionized gas. The Br$\gamma$ line flux
  is comparable to the flux of H$_{2}$ at 1-0 S(1) transition. \citet{gledhil15} carried out 
  a detailed study of the 2\,$\mu$m to 2.4\,$\mu$m spectrum of this star and found that the Br$\gamma$ 
  emission is spatially extended tracing the extended ionized hydrogen region. We also detected He\,{\sc i} 
  line at 2.058\,$\mu$m.  We find from our spectrum
  that the H$_{2}$ line flux ratios 1-0 S(1)/2-1 S(1) = 3.62 indicating that the overall excitation of
  H$_{2}$ in this object is due to UV fluorescence. This is in agreement with the suggestion by \citet{kelly15}  
   in their H$_{2}$ 2-micron spectral survey of post AGB objects (the ratio given by
  them is 1.5). However, the line flux ratio derived later by \citet{gledhil15} is 5\,$\pm$\,2 and argued based on morphology
  that most of the H$_{2}$ had been disassociated.  \citet{gledhil15} showed two clumps of 1-0 S(1) emission plus what looked like
  faint emission across the 4$''$$\times$ 4$''$ field. \citet{kelly15} showed emission extending 12$''$ along the 2.4$''$ 
  wide E-W slit. Their fluxes were probably dominated by the extended emission which one would expect to be excited radiatively. 
  Our observations could easily be different region than were surveyed in the  two papers mentioned above. \citet{garcia97} and \citet{arkh01} made a detailed
  study of the optical spectrum of this rapidly evolving hot post-AGB star  and found that absorption and emission profiles are variable.  This star is evolving
  rapidly to the young planetary nebula phase \citep{partha93b,1994ASPC...60..261P}.\\ 
   
 \subsection{IRAS\,20056+1834}

  The late type variable star IRAS\,20056+1834 has a spectral type of G0e. We obtained the K-band spectrum 
 of the star for the first time. No H$_{2}$ lines and Br$\gamma$ line are detected in our spectrum (Figure 1). 
 However, it shows CO (2-0) and CO (3-1) absorption lines at 2.29\,$\mu$m and 2.32\,$\mu$m respectively. \citet{menz98}
  discussed the optical spectrum of this star and later, \citet{kloch07} analyzed repeated high resolution optical spectrum of this star and found that the photospheric 
 lines are variable. Their chemical composition study of the star shows that it is overabundant in carbon and nitrogen.\\ 

 \subsection{IRAS\,19399+2312 (V450 Vul)}
 
 IRAS\,19399+2312 is a B0IVe type star. It may not be a post-AGB star (see \citet{2009ApJ...703..585C}) however,
 it is possible that it could be a hot subdwarf. Our K-band spectrum does not show either
 H$_{2}$, He\,{\sc i} or Br$\gamma$ emission lines (Figure 1). The CO absorption lines at 2.2935\,$\mu$m and 2.322\,$\mu$m seem to be present. \\

 \subsection{IRAS\,19475+3119 (HD\,331319)}

 The post AGB star IRAS\,19475+3119 is of F3Ibe spectral type \citep{partha93a}. It shows a 
 quadrupolar nebular morphology \citep{sahai07}. \citet{hriv94} have done near-IR spectral
 study of this cool post AGB star and found Br$\gamma$ in absorption. Our K-band spectrum indicates a
 broad Br$\gamma$ absorption profile with possible weak emission in the wings (Figure 1). No H$_{2}$ lines are
 seen. \citeauthor{kelly15} has included this source in their 
 H$_{2}$ emission survey. They also did not find H$_{2}$ line. \citet{siv01} analyzed high resolution
 optical spectra of this post-AGB star and derived the stellar atmospheric parameters and found that
 abundances are due to mixing of third dredge-up, and nitrogen is overabundant. \\

 \subsection{IRAS\,23304+6147}

 \citet{kloch00} analyzed high resolution optical spectrum of this post AGB star. They
 find the photospheric temperature T$_{eff}$ of 5900 K and log g = 0.0. They find this object
 to be a metal poor (Fe/H = $-$0.6) and show over abundance of carbon and s-process elements. \citet{kelly15}  
 included this object in their 2\,$\mu$m spectral survey and found Br$\gamma$ in
 absorption. They found the spectral class of this object to be G2 Ia. We find from our K-band 
 spectrum of this post-AGB star that the Br$\gamma$ is a broad absorption feature (Figure 1). We do not see 
 any H$_{2}$ lines or He {\sc i} line in our spectrum. \\

 \subsection{IRAS\,05040+4820 (SAO\,40039)} 
 
 \citet{partha93a} classified IRAS\,05040+4820 as a post-AGB supergiant based on its IRAS 
 colours and flux distribution. \citet{Fujii02} made $B,V,R,I,J,H,K$ photometry and studied the 
 flux distribution and derived the circumstellar dust shell parameters and concluded that it is a 
 low mass post-AGB star. Its spectral type is A4Ia. \citet{partha05}  made $B,V,R,I$
 polarization observations and concluded that the circumstellar dust shell has asymmetric shape. 
 Our K-band spectrum of this post-AGB star shows that the Br$\gamma$ is a broad absorption line (Figure 1). 
 No H${_2}$ lines or He\,{\sc i} line are seen in our K-band spectrum for this object. \\
 
\section{Discussion and Conclusions}

 K-band (2\,$\mu$m to 2.4\,$\mu$m) spectra of 12 post-AGB candidates are presented. In our sample
 seven stars are hot post-AGB stars. Seven objects 
 are found to show Br$\gamma$ emission lines at 2.166\,$\mu$m implying ionized hydrogen region in the
 nebulae. Three sources show Br$\gamma$ in absorption and three have Br$\gamma$ P-Cygni profile
 indicating ongoing mass-loss in these objects. Four sources show He\,{\sc i} line at 2.058\,$\mu$m. H$_{2}$ 2.1218\,$\mu$m emission line is present in the
 spectra of  IRAS\,06556+1623, IRAS\,22023+5249, IRAS\,18062+2410 (SAO 85766) and IRAS\,20462+3416. These four sources
 also were seen to show  variation in Br$\gamma$ and H$_{2}$ emission line strengths when compared 
 with other studies mentioned above. Four of the sources show H$_{2}$ transitions at 1-0 S(1) and
 2-1 S(1) (Figure 1, Table 1). The H$_{2}$ flux ratio 1-0 S(1)/2-1 S(1) indicates that the dominant excitation of H$_{2}$ 
 is by UV florescence. The hot post-AGB stars 
 IRAS\,22495+5134, IRAS\,22023+5249, IRAS\,18062+2410 and IRAS\,20462+3416 seem to be evolving rapidly to 
 the planetary nebula phase. Further monitoring of stars showing Br$\gamma$ and H$_{2}$ emission lines is important.
 Some of the cooler stars in which the Br$\gamma$ line is in absorption is most likely stellar photosperic origin. \citet{marq13}
 conclude that H$_{2}$ emission is not exclusive of bipolar planetary nebulae (PNe) and proto-planetary nebulae
 (PPNe), although objects with this morphology are still the brightest H$_{2}$ emitters.

The study of \citet{gledhil15} revealed the spatial extent
of the emission line regions in IRAS\,18062+2410 (SAO\,85766), IRAS\,20462+3426 (LS II +34 26),
and IRAS\,22023+5249 (LS III +52 24). In our study and in our data reduction we have not
made an attempt to estimate the expected spatial extent of the sources and how
it affected the fluxes measured through the 2.4$''$ width E-W slit.

The rapidly evolving hot post-AGB stars mentioned above are of significant
importance as these are rare objects and the evolution of their circumstellar envelopes
and central stars can be studied in real time. IRAS\,18062+2410 (SAO\,85766)
which is observed by all four of this studies mentioned above and also by \citet{2003MNRAS.344..262D} (observed in June 2000)
evolved from a 8500\,K post-AGB star to 20,000\,K post-AGB star in 20\,years \citep{partha00b},
and its ionized mass increased by a factor of three in a short span of 8\,years \citep{2011MNRAS.412.1137C}.
The Br$\gamma$ and radio flux increased linearly indicating that the ionization of the circumstellar envelope started
around 1990 \citep[see][]{gledhil15}. IRAS\,20462+3426, and IRAS\,22023+5249 are in similar
rapid phase of evolution of their cental stars and their circumstellar envelopes. Further K-band
spectroscopic monitoring of these rapidly evolving post-AGB objects is needed.

\acknowledgments

  MP is thankful to Dr. T. Fujii for his help during the observations. MP is thankful to the Director 
General of NAOJ and Prof. Y. Nakada and Prof. S. Deguchi for their kind encouragement and support. We are thankful to the referee
for helpful comments.

\section{Conflict of Interest}
The authors declare that they have no conflict of interest.

\clearpage
 \begin{figure}
\plotone{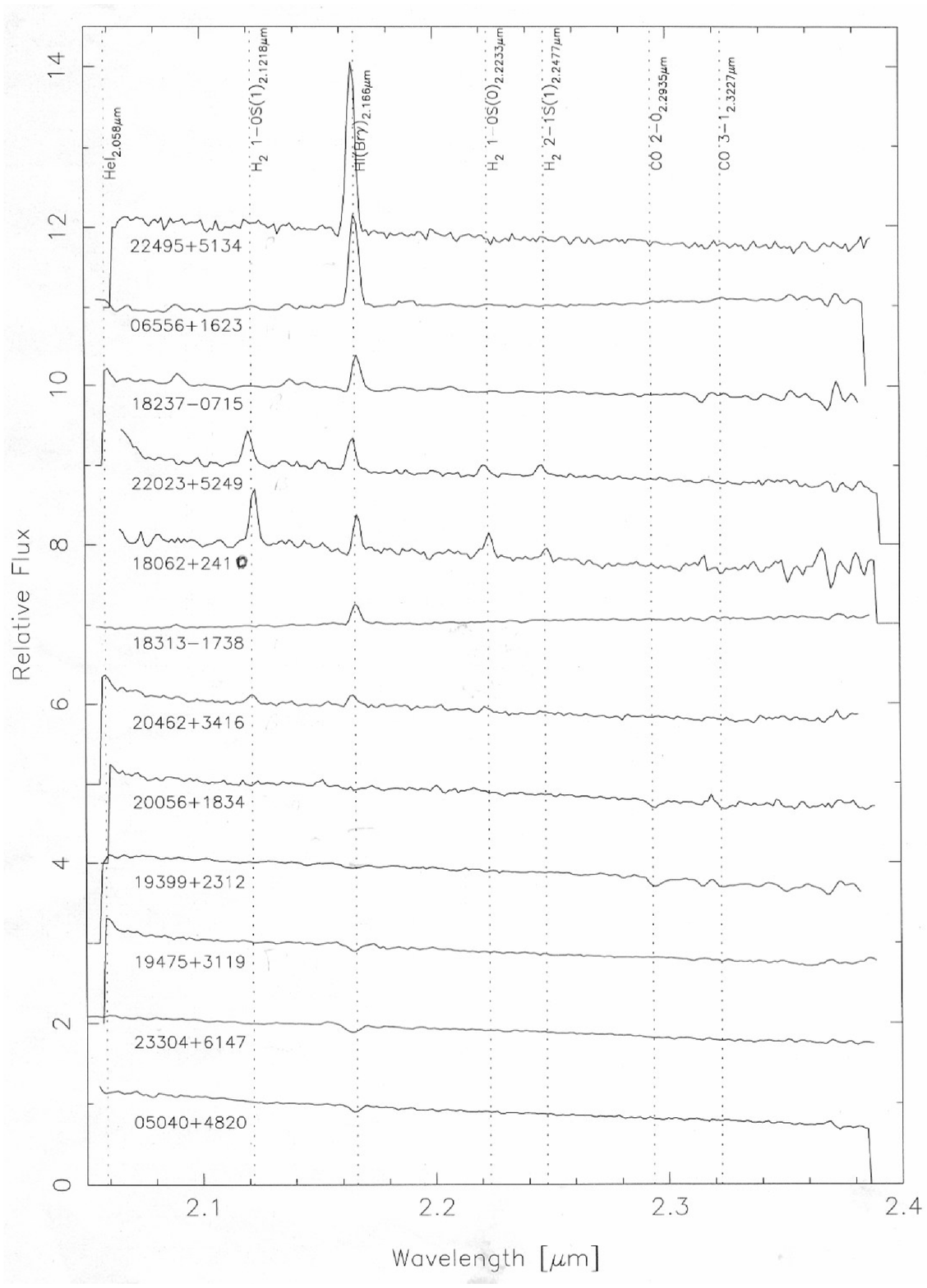}
\caption{K-band spectra of selected post-AGB candidates.\label{fig1}}
\end{figure}

\clearpage

\begin{figure}
\plotone{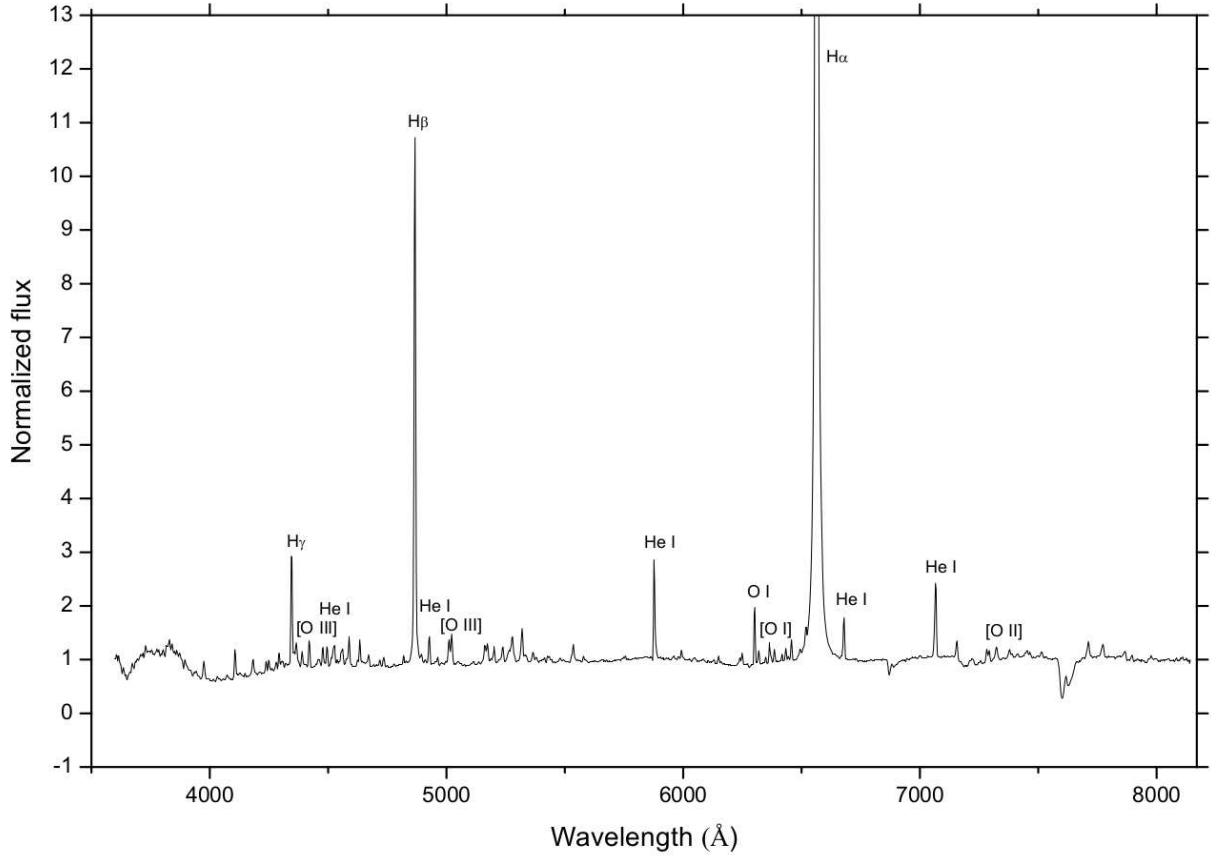}
\caption{Medium resolution spectrum of IRAS 06556+1623.\label{fig2}}
\end{figure}

\clearpage

\begin{landscape}
\begin{table}
\begin{center}
\caption{Relative Line fluxes \label{tbl2}}
\vspace{.5cm}
\begin{tabular}{lccccccc}
\tableline\tableline
           &        &      &      &     &     &   \\
Object & Br$\gamma$ & H$_{2}$ 1-0 S(1) & H$_{2}$ 2-1 S(1) & H$_{2}$ 1-0 S(0) & He {\sc i} & 1-0 S(1)/2-1 S(1) & \\
IRAS ID & 2.166\,$\mu$m & 2.1218\,$\mu$m & 2.2477\,$\mu$m & 2.2233\,$\mu$m & 2.058\,$\mu$m    & & Sp. Type\\
           &        &      &      &     &     &  & \\
\tableline
           &        &      &      &     &     &  & \\
22495+5134 & 93.69 & -- & -- & -- & -- & --& B0e\\
06556+1623 & 55.69 & 2.27 & 0.48 & 0.93 & Yes & 4.73& BeI/BQ[\,]\\
18237-0715 & 18.90 & --  & --   & --  & Yes  & --& B5/9Iaeq\\
22023+5249 & 17.84 & 20.69 & 6.85 & 6.62 & No & 3.02& B0e\\
18062+2410 & 17.24 & 27.18 & 5.56 & 13.48 & -- & 4.89& B1IIIe\\
18313-1738 & 11.84 & -- & -- & -- & -- & No    --& Be\\
20462+3416 & 6.78  & 5.58 & 1.54 & 2.49 & Yes  & 3.62 & B1.5Ia/Iabe\\
20056+1834 &  --   &  --  & --   &  --  & No &  --& G0Ie\\ 
19399+2312 &   --   &  --  & --   &  -- & No &  --& B0IVe\\
19475+3119 & absorp &  --  & --   &  -- & Yes & --& F3Ibe\\
23304+6147 & absorp &  --  & --   &  -- & No  & --& G2Ia\\
05040+4820 & absorp &  --  & --   &  -- & No  & --& A4Ia\\
           &        &      &      &     &     &   & \\
\tableline
\end{tabular}
\end{center}
\end{table} 
\end{landscape}


\begin{thebibliography}{}

\bibitem[Arkhipova et al. (2001)]{arkh01} Arkhipova, V. P., et al., 2001, Ast.Lett, 27, 719 
\bibitem[Arkhipova et al. (2006)]{arkh06} Arkhipova, V. P., et al., 2006, Ast.Lett, 32, 594
\bibitem[Arkhipova et al. (2013)]{arkh13} Arkhipova, V. P., et al., 2013, Ast.Lett, 39, 619
\bibitem[Cerrigone et al.(2009)]{2009ApJ...703..585C} Cerrigone, L., Hora, J.~L., Umana, G., et al.\ 2009, \apj, 703, 585
\bibitem[Cerrigone et al.(2011)]{2011MNRAS.412.1137C} Cerrigone, L., Trigilio, C., Umana, G., et al.\ 2011, \mnras, 412, 1137 
\bibitem[Davis et al.(2003)]{2003MNRAS.344..262D} Davis, C.~J., Smith, M.~D., Stern, L., et al.\ 2003, \mnras, 344, 262
\bibitem[Black \& Dalgarno (1976)]{black76} Black J.H. \& Dalgarno A. 1976, ApJ, 303, 132
\bibitem[Fujii et al. (2002)]{Fujii02} Fujii, T, Nakada, Y., Parthasarathy, M., 2002, \aap, 385, 884, 
\bibitem[Garcia-Hernandez et al. (2002)]{garcia02} Garcia-Hernandez, D. A., et al., 2002, \aap, 387, 955
\bibitem[Garcia-Lario et al. (1997)]{garcia97} Garcia-Lario, P., Parthasarathy, M., de Martino, D., et al., 1997, \aap, 326, 1103 
\bibitem[Gauba \& Parthasarathy (2003)]{gauba03} Gauba, G., \& Parthasarathy, M, 2003, \aap, 407, 1007
\bibitem[Gauba \& Parthasarathy (2004)]{gauba04} Gauba, G., \&  Parthasarathy, M, 2004, \aap, 417, 201
\bibitem[Gledhill \& Forde (2015)]{gledhil15} Gledhill, T. M., \& Forde, K. P., 2015, \mnras, 447, 1080
\bibitem[Handler(1999)]{1999A&AS..135..493H} Handler, G.\ 1999, \aaps, 135, 493
\bibitem[Hrivnak et al. (1994)]{hriv94} Hrivnak, B.J., Kwok, S.,  Geballe, T. R., 1994, \apj, 420, 783
\bibitem[Jaschek et al.(1996)]{1996A&AS..117..281J} Jaschek, C., Andrillat, Y., \& Jaschek, M.\ 1996, \aaps, 117, 281
\bibitem[Kelly \& Hrivnak (2005)]{kelly15} Kelly, D. M., \& Hrivnak, B. J., 2005, \apj, 629, 1040
\bibitem[Klochkova et al. (2007)]{kloch07} Klochkova, V. G., et al., 2007, AstBu, 62, 217 
\bibitem[Klochkova et al. (2000)]{kloch00} Klochkova, V. G., et al., 2000, Ast.Lett, 26, 88
\bibitem[Marquez-Lugo et al. (2013)]{marq13} Marquez-Lugo, R., et al., 2013, \mnras, 429, 973
\bibitem[Miroshnichenko et al. (2005)]{miro05} Miroshnichenko A. S., et al., 2005, \mnras, 364, 335
\bibitem[Menzies \& Whitelock (1998)]{menz98} Menzies, J.W., \& Whitelock, P. A., 1988, \mnras, 233, 697 
\bibitem[Okamura et al. (2000)]{oka2000} Okamura, S., et al., 2000, \pasj, 52, 931
\bibitem[Olivia \& Origlia (1992)]{oli92} Olivia, D., \& Origlia, L., 1992, \aap, 254, 266
\bibitem[Oudmaijer et al.(1995)]{1995A&A...299...69O} Oudmaijer, R.~D., Waters, L.~B.~F.~M., van der Veen, W.~E.~C.~J., et al.\ 1995, \aap, 299, 69
\bibitem[Parthasarathy \& Pottasch(1986)]{1986A&A...154L..16P} Parthasarathy, M., \& Pottasch, S.~R.\ 1986, \aap, 154, L16
\bibitem[Parthasarathy \& Pottasch(1989)]{1989A&A...225..521P} Parthasarathy, M., \& Pottasch, S.~R.\ 1989, \aap, 225, 521
\bibitem[Parthasarathy (1993a)]{partha93a} Parthasarathy, M., 1993a, ASP Conference Series, 45, 173 
\bibitem[Parthasarathy (1993b)]{partha93b} Parthasarathy, M., 1993b, \apj, 414, L109
\bibitem[Parthasarathy (1994)]{1994ASPC...60..261P} Parthasarathy, M.\ 1994, \emph{The MK Process at 50 Years: A Powerful Tool for Astrophysical Insight}, ASP Conference Series, Vol. 60, 261
\bibitem[Parthasarathy et al. (1993)]{partha93} Parthasarathy, M., et al., 1993, \aap, 267, L19 
\bibitem[Parthasarathy et al. (1995)]{partha95} Parthasarathy, M., et al., 1995, \aap, 300, L25
\bibitem[Parthasarathy et al. (2000a)]{partha00a} Parthasarathy, M., Vijapurkar, J., Drilling, J. S., 2000a, \aap, 145, 269
\bibitem[Parthasarathy et al. (2000b)]{partha00b} Parthasarathy, M., Garcia-Lario, P., Sivarani, T., et al., 2000b, \aap, 357, 241
\bibitem[Parthasarathy et al. (2005)]{partha05} Parthasarathy, M., Jain, S. K., Sarkar, G., 2005, \aj, 129, 2451
\bibitem[Sahai et al. (2007)]{sahai07} Sahai, R., et al., 2007, \apj, 658, 410
\bibitem[Sarkar et al. (2012)]{sark12} Sarkar, G., Garcia-Hernandez, D. A., Parthasarathy, M., et al., 2012, \mnras, 421, 679
\bibitem[Sivarani et al. (2001)]{siv01} Sivarani, T., Parthasarathy, M., Garcia-Lario, P., et al., 2001, ASSL, 265, 301
\bibitem[Smith (1995)]{smi95} Smith M.D. 1995, A\&A, 296, 789
\bibitem[Vijapurkar et al. (1998)]{vijap98} Vijapurkar, J., Parthasarathy, M., Drilling, J. S., 1998, BASI, 26, 497

\end{thebibliography}
\end{document}